\RequirePackage{fix-cm}
\documentclass[smallextended]{svjour3}
\smartqed  % flush right qed marks, e.g. at end of proof

\usepackage{graphicx}

\usepackage[outdir=./figuras/]{epstopdf}
\usepackage[normalem]{ulem}
\usepackage{lineno,hyperref}
\usepackage{amsmath,amssymb,amsfonts}
\usepackage{algorithmic}
\usepackage{wrapfig}
\usepackage{textcomp}
\usepackage{color, colortbl}
\usepackage{multirow}
\usepackage{flushend}
\usepackage{longtable}
\usepackage{booktabs}
\usepackage[table,xcdraw]{xcolor}
\usepackage{natbib}
\usepackage[utf8]{inputenc}
\usepackage{lscape}
\usepackage{booktabs}
\usepackage[utf8]{inputenc}
\usepackage{todonotes}
\usepackage{natbib}
\usepackage[linesnumbered,ruled]{algorithm2e}
\usepackage{enumitem}
\usepackage{nomencl}
\usepackage{subfigure}
\usepackage[flushleft]{threeparttable}
\usepackage{color, colortbl}
\usepackage{scalefnt}
\usepackage{adjustbox}
\usepackage{framed}
\usepackage{float}
\usepackage{xspace}
\usepackage{comment}
\usepackage{float}
\usepackage{verbatim}
\usepackage{listings}
\usepackage{siunitx,booktabs}
\usepackage{graphicx}

\newcommand{\totalMultipleLicenses}{947\xspace}
\newcommand{\totalLicenses}{259,214\xspace}
\newcommand{\totalUniqueLicenses}{434\xspace}
\newcommand{\totalUniqueNonSoftwareLicenses}{74\xspace}
\newcommand{\totalProjects}{1,552\xspace}
\newcommand{\totalFiles}{1,426,263\xspace}
\newcommand{\totalProjectsAnaliseManual}{776\xspace}

\newcommand{\zenodo}{https://zenodo.org/record/3609284}

\newcommand{\projectsWithLicenseIssues}{373\xspace}
\newcommand{\totalSurvey}{83\xspace}
\newcommand{\totalFirstSurveyInvitation}{41\xspace}
\newcommand{\totalSecondSurveyInvitation}{42\xspace}
\newcommand{\totalInvitedSurvey}{1,366\xspace}

\definecolor{Gray}{gray}{0.9}

\newcolumntype{L}[1]{>{\raggedright\let\newline\\\arraybackslash\hspace{0pt}}m{#1}}
\newcolumntype{C}[1]{>{\centering\let\newline\\\arraybackslash\hspace{0pt}}m{#1}}
\newcolumntype{R}[1]{>{\raggedleft\let\newline\\\arraybackslash\hspace{0pt}}m{#1}}

\newcommand{\MyBox}[1]{\vspace{3mm}\noindent\framebox[\columnwidth][c]{\parbox[b]{0.95\columnwidth}{ #1 }}\vspace{.2em}}

\begin{document}

\epstopdfsetup{outdir=./figuras/}

%\title{How and Why do JavaScript Projects Inherit Multiple Open Source Licenses?}

\title{From One to Hundreds: \\ Multi-Licensing in the JavaScript Ecosystem}

\author{João Pedro Moraes \and
        Ivanilton Polato \and
        Igor Wiese \and
        Filipe Saraiva \and \\
        Gustavo Pinto
}

\institute{João Pedro Moraes \and  Gustavo Pinto
\at Federal University of Par\'a (UFPA), Brazil \\
\email{gpinto@ufpa.br}
\and
Filipe Saraiva
\at Federal University of Par\'a (UFPA), Brazil \\
KDE e.V.\\
\email{saraiva@utfpr.edu.br, filipe@kde.org}
\and
Ivanilton Polato \and Igor Wiese
\at Federal Technological University of Paran\'a (UTFPR), Brazil \\
\email{\{igor, ipolato\}@utfpr.edu.br}
}

\date{Received: date / Accepted: date}

\maketitle

\begin{abstract}

Open source licenses create a legal framework that plays a crucial role in the widespread adoption of open source projects. Without a license, any source code available on the internet could not be openly (re)distributed.
Although recent studies provide evidence that most popular open source projects have a license, developers might lack confidence or expertise when they need to combine software licenses, leading to a mistaken project license unification.
This license usage is challenged by the high degree of reuse that occurs in the heart of modern software development practices, in which third-party libraries and frameworks are easily and quickly integrated into a software codebase.
This scenario creates what we call ``multi-licensed'' projects, which happens when one project has components that are licensed under more than one license. Although these components exist at the file-level, they naturally impact licensing decisions at the project-level.

In this paper, we conducted a mix-method study to shed some light on these questions. We started by parsing \totalFiles (source code and non-source code) files available on \totalProjects JavaScript projects, looking for license information. Among these projects, we observed that \totalMultipleLicenses projects (61\%) employ more than one license. On average, there are 4.7 licenses per studied project (max: 256).
Among the reasons for multi-licensing is to incorporate the source code of third-party libraries into the project's codebase.
When doing so, we observed that \projectsWithLicenseIssues of the multi-licensed projects introduced at least one license incompatibility issue.
We also surveyed with \totalSurvey maintainers of these projects aimed to cross-validate our findings. We observed that 63\% of the surveyed maintainers are not aware of the multi-licensing implications. For those that are aware, they adopt multiple licenses mostly to conform with third-party libraries' licenses.

\keywords{Open Source Licenses \and Multi-licensing \and JavaScript projects}

\end{abstract}

%\linenumbers

\section{Introduction}
\label{sec:introduction}

An open source license (or simply license, for short) grants developers the freedom to use, reproduce, change, and redistribute the source code.
Open source projects must then always be distributed under an open source software license. Without a license, all these rights are then restricted to the project's owner, limiting its general use. It means that although a source code is available on source coding platforms such as GitHub or GitLab, that source code is not open source and, thus, could not be redistributed. Therefore, a software license is imperative to open source rise and thrive~\citep{Wu:2017:EMSE}.

Even though most of the popular open source projects do have a license~\citep{Vendome:2017:ESME,Vendome:2015:ICPC}, due to its non-technical nature, developers still struggle to choose the appropriate license for their software projects~\citep{Almeida:2017:ICPC}.
Moreover, developers might not feel completely confident or might lack enough expertise when combining software licenses~\citep{Almeida:2017:ICPC}. %This can lead to a mistaken project license unification when developers choose a single open source license while they should be considering all the licenses they are overlaying.
%\sout{there is a recurring belief that open source projects should be licensed using a single open source license.}
This concern is exacerbated by the high degree of software reuse in the heart of modern software development practices.

Not so long ago, developers that relied on software stacks such as LAMP (i.e., Linux, Apache, MySQL, and PHP), would have to care only about a handful of licenses. However, due to the advances in software development techniques, including package manager systems~\citep{McIntosh:2012:EMSE} and social coding websites~\citep{Storey:2017:TSE}, developers could straightforwardly find and reuse third-party libraries or frameworks that fix their problems with ease~\citep{Campos:MSR:2019,Zhang:ICSE:2018}, eventually helping them create their own development stack.

This process of integrating third-party libraries into their codebase might introduce risks to software licensing, especially when combining licenses from different classes, such as permissive and copyleft licenses~\citep{Kapitsaki:JSS:2017,Kapitsaki:JSS:2015}.
This concern is particularly relevant to the JavaScript ecosystem, since 1) JavaScript libraries are readily available (either by source code or packages through its well-known package manager, NPM), 2) this availability of JavaScript libraries promotes a high degree of software reuse in JavaScript projects, and 3) which might ended up creating a complex relationship of license usage in this ecosystem.

%For instance, if a developer wants to use the \textsf{GateOne} library, a HTML5 terminal emulator licensed under AGPL, the developer's project should be license as AGPL as well (one of the restrictions of the GPL family of licenses). \gnote{confirmar essa restrição da AGPL} This problem is exacerbated when developers reuse several third-party libraries, creating

One license usage dimension that was not yet fully explored is what we call \emph{multi-licensing}, that is, when one open source project has multiple internal components (e.g., modules, classes, files, etc) that are licensed under different licenses.
Take the \textsf{three.js} project\footnote{Available at \url{https://github.com/mrdoob/three.js}}---a very popular JavaScript library (63k stars and 25k forks) for creating 3D web applications---as a concrete example. Among many other source code files, this project has a source code file named \texttt{VolumeShader.js}, licensed under MIT license, another source code file named \texttt{esprima.js}, licensed under BSD-2, and yet another source code file named \texttt{jszip.min.js}, licensed under GPL-3.0. Consequently, the \textsf{three.js} maintainers should take this information into account when deciding how to license \textsf{three.js}. Unfortunately, maintainers have to rely on their own licensing skills, since tool support is far from mature, and current social coding websites or package management systems provide little guidance.

To better understand the current landscape of multi-licensing in open source projects, we conducted a mix-method study. We started by mining and parsing
\totalFiles files from \totalProjects popular JavaScript open source projects to understand their multi-license usage. We relied on ScanCode Toolkit\footnote{Available in: \url{https://github.com/nexB/scancode-toolkit}} (ScanCode for short), a top-notch industry tool for inferring open source licenses, to conduct this task. %Overall we noticed that 62\% of the studied projects employ more than one license.
We then surveyed \totalSurvey maintainers to understand if they take multi-licensing into account when licensing their projects. Our main findings are in the following:

\begin{itemize}
    \item We found a total of \totalMultipleLicenses (out of the \totalProjects, 62\%) JavaScript projects licensed under more than one license. Many of these projects use multiple licenses because they include other JavaScript projects (as minified versions) in their sub-directories.

    \item We discovered that license incompatibilities exist in \projectsWithLicenseIssues multi-licensed projects. These license incompatibilities happen when a project uses a source code file that is licensed under a non-compatible license with the license of the main repository. For instance, 309 of these projects are licensed under MIT, but uses source code files licensed under more restrictive licenses.

    \item We observed that, when considering the pair of licenses frequently used together, the most common pair of licenses are the ``BSD-3 and MIT'' pair, followed by the ``Apache-2.0 and MIT'' pair. Both pairs are based on permissive licenses.

    \item We noted that 17\% of the projects adopted at least one non-software license. The OFL license is, by far, the most common one. When considering the pair of (software and non-software)  licenses, OFL is often employed along with MIT, Apache-2.0 and BSD-3.

    \item We observed that developers adopt multiple licenses to avoid problems with license compatibility, legal third-party library adoption, and to impose fewer restrictions on their source code.
\end{itemize}

%\gnote{revisar - added (wiese)}

\section{Definitions}

Open source software licensing has many, sometimes unclear and conflicting, definitions. In this section we briefly revisit the main characteristics of open source license. For a comprehensive guide, we recommend reading the book of~\cite{meeker2017open}.

The fundamental principle in open source is that the source code is available; if the source code is not clearly available, there is no open source. To permit others to use the source code, the copyright owner, that is, the developer who wrote that code (and holds its copyright), should attach a license that gives permissions to use the source code. An open source license is a legal mechanism that specify how users should use the source code. Any open source project should be accompanied by an open source license. Without a license, users do not have permission to use the source code (either in part or complete).

While some open source licenses are more flexible, allowing a program that is once open source to become closed source (i.e., proprietary) at any moment in the future, some other open source licenses make sure that a program that is once open source will always be open source. There are three leading groups of licenses: permissive, copyleft, and weak-copyleft.

A \emph{permissive} license employs minimal requirements about how the software could be redistributed. Permissive licenses do not place many restrictions on derivative works, meaning that an open source software licensed under a permissive license can later become ``closed''. Notable copyleft licenses include: the BSD license, the MIT license, and the Apache license.

A \emph{copyleft} license guarantees that a program licensed under such a license is free to use, run, study, and redistribute without permissions. Those that advocate in favor of copyleft licenses argue that software should be free for everyone. Therefore, to benefit a broader population, their derivative works should remain free as well. While copyright restricts the way users could use a program (e.g., there is no redistribution), copyleft, on the other hand, reinforces the freedom to exploit a program for any purposes. Notable copyleft licenses include: the GPL license and the AGPL license.

There is also \emph{weak copyleft}, which is a family of licenses that shares some copyleft ideas and are considered to be in the middle ground between copyleft licenses and permissive licenses. As an example, a program licensed under Mozilla Public License (MPL) can either be converted to a copylefted program or to a proprietary source program (as long as the source code is available under MPL). Notable copyleft licenses include: the MPL license and the EPL license.

%\fnote{\textbf{REESCREVENDO BASEADO NO TXT DO GUSTAVO ABAIXO - SE ACEITO REMOVER O DO GUSTAVO}

\begin{figure}[H]
    \centering
    $
    \begin{array}{cc}
        \frame{\includegraphics[scale=0.55]{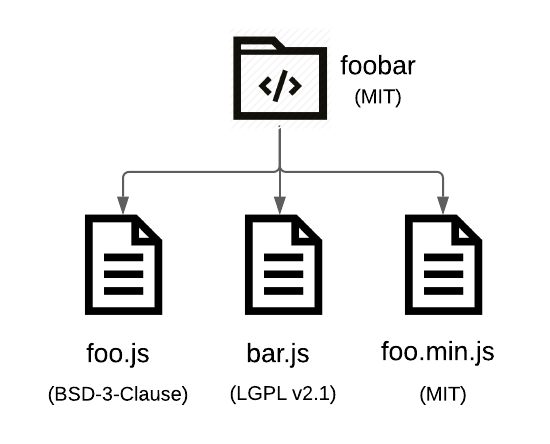}} & \frame{\includegraphics[scale=0.47]{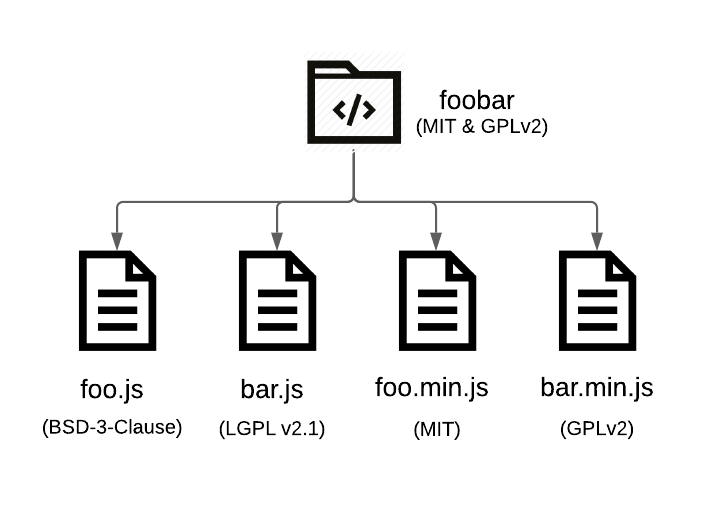}} \\
        \textsf{(a)} & \textsf{(b)}  \\
    \end{array}
    $
    \caption{Multi-licensing in a nutshell. Figure (a) shows multi-licensing at the file-level, while figure (b) complements this vision, but now considering multi-licensing at the project-level.}
    \label{fig:multi-licensing}
\end{figure}

In the context of this work, we talk about \emph{multi-licensed} open source projects. This kind of license usage happens when one project has components (i.e., a library, a module, a file, etc.) that are licensed under different licenses. Figure~\ref{fig:multi-licensing}-(a) shows an illustrative example of a multi-licensed project. In this illustrative example, our project \textsf{foobar} has three files: \texttt{foo.js}, \texttt{bar.js}, and \texttt{foo.min.js}, which are licensed under, respectively, BSD, LGPL, and MIT license. In this particular example, licensing \textsf{foobar} under MIT was an appropriate choice because MIT is compatible with BSD and LGPL\footnote{We could license \textsf{foobar} under MIT if the \texttt{bar.js} file is not changed. If that file is changed, the project has to be licensed under LGPL.}. %We call this as \textbf{multi-licensing at the file-level}.

Consider now our extended example, in Figure~\ref{fig:multi-licensing}-(b). In this example, the \textsf{foobar} project has one additional file, \texttt{bar.min.js}, which is licensed under GPL-2.0. In this case, the MIT license cannot cover all the restrictions imposed by GPL-2.0. An eventual solution to this problem (as also reported by one of our survey respondents, in Section~\ref{sec:rq5}) is to re-license \textsf{foobar} under MIT \emph{and} GPL-2.0. When this is needed, we call this action as \textbf{multi-licensing at the project-level}.

Throughout this paper, we refer to file-level and project-level definitions to present and discuss our results.

%Regardless if on the project-level or on the file-level, we argue that open source maintainers must pay attention to every new component they introduce into their programs' codebase.

%The multi-licensing phenomenon is amplified due to the presence of package managers and social coding environments, which makes software reuse rather straightforward.

%Notice, however, that a multi-licensed project does not imply that the project is licensed under more than one license; this could indeed be another perspective of multi-licensing. However, we do not explore this dimension in this work.

\section{Research Questions}\label{chap:metodologia}

This work is guided by the following research questions.

\begin{itemize}
    \item[\textbf{RQ1.}] How common are multi-licensed JavaScript projects?
\end{itemize}

\noindent
\textbf{Rationale.} JavaScript is a very popular programming language with a vivid community that produces dozens of new libraries and frameworks on a regular basis. JavaScript developers are also regarded as very active when it comes to software reuse, which is particularly due to the culture of absorbing the new third-party libraries that are introduced in the software market regularly. This culture poses a challenge, though. Since one simple JavaScript application often depends on other several packages (on average, JavaScript projects published on the NPM website depend on $\sim$5 other packages~\cite{Meloca:2018:MSR}), developers of this community should place additional care in filtering out packages that do not offer to them license compliance.%; or if they are really into adding new packages, they may do this at the cost of extending their licenses.
In this research question, we investigate how common is multi-licensing in JavaScript projects.

\begin{itemize}
    \item[\textbf{RQ2.}] What are the pair of licenses most used together?
\end{itemize}

\noindent
\textbf{Rationale.} Given that JavaScript projects using more than one license in their internal codebase might be commonplace, it would also be interesting to investigate the pair of licenses that are commonly employed together. A license combination may be a difficult task and can introduce legal violations during the process. Therefore, the investigation of the most common licenses applied together can clarify these issues to developers.
We grouped the licenses into two groups to better understand this behavior: software licenses and non-software licenses.
Unlike software licenses, non-software licenses are used on artifacts other than source code, such as documentation, pictures, music, etc. Our corpus of open source projects may also take advantage of this kind of artifact (e.g., for a web-based application, it is relatively common to use images or icons).

\begin{itemize}
    \item[\textbf{RQ3.}] How common are license incompatibility issues in multi-licensed projects?
\end{itemize}

\noindent
\textbf{Rationale.} Understanding whether one license is compatible with another is a complicated endeavor since there are dozens of approved licenses (and hundred other known licenses).
According to Kapitsaki, ``license compatibility refers to the ability of combining different FOSS licenses into the same software product'' ~\citep{Kapitsaki:JSS:2017}.
In this research question, we used the compatibility graph introduced by~\cite{Kapitsaki:JSS:2017} to assess the presence of license compatibility issues in the studied projects. %This graph only focuses on a subset of the approved open source licenses.

\begin{itemize}
    \item[\textbf{RQ4.}] Are multi-licensed projects warning their potential clients on the use of multiple licenses?
\end{itemize}

\noindent
\textbf{Rationale.} When developers are looking for libraries that they can use on their projects, they have to make sure that the library's license conforms with the license that their original project employs. To ease such navigation, developers often keep an eye open to traditional files, such as the \texttt{LICENSE} file or the \texttt{README} file.
However, when adopting multiple licenses, maintainers of these libraries must have to update several ``traditional files'' to make their clients aware of their license usage. This concern is particularly relevant for web infrastructures that made the license information clearly available. For instance, GitHub parses the \texttt{LICENSE} file to infer the license used, whereas NPM (the main JavaScript package manager) reads the information available on the \texttt{package.json} file. In this research question, we sought to investigate the extent to which these projects keep their license information clearly available for potential clients.

\begin{itemize}
    \item[\textbf{RQ5.}] Are developers aware of multi-licensing issues in their projects?
    %Why do developers adopt multiple open source licenses?
\end{itemize}

\noindent
\textbf{Rationale.} In this final research question, we are interested in understanding if developers understand the issues when multi-licensing a project. We surveyed developers that modified license-related files since they could be more aware of licensing decisions.
We also collected 927 commits that modified license-related files. We randomly sampled 195 commits (confidence level of 95\% with a ±5\% error) on these license files to conduct a manual analysis.
We hypothesized that the changes that occur on license-related files could help to pile additional evidence on the reasons for adopting multiple licenses.

\section{Research Approach}

To help us answer our research questions, we conducted two different studies. The first one is a mining software repositories study (Section~\ref{sec:mps}), while the second study performs a survey with open source developers (Section~\ref{sec:survey}).

\subsection{Study 1: A Software Repository Mining}\label{sec:mps}

This section describes the steps employed to select the studied projects (Section~\ref{selecaoProjetos}) and the procedure used to inspect them (Section~\ref{analiseDados2}).

\subsubsection{Selecting JavaScript Projects}
\label{selecaoProjetos}

To curate the set of studied projects, we used the dataset provided by \cite{Borges:ICSME:2016}, which collected the 5,000 most popular open source repositories hosted on GitHub (sorted by the number of stars, as of January 2017). This dataset is made  available in a CSV format, in the Zenodo platform\footnote{Available in: \texttt{\url{https://zenodo.org/record/804474}}}. However, in this work, we focused only on JavaScript projects, which reduced the number of projects to \totalProjects. Our motivation to focus solely on JavaScript projects is based on at least three important observations:

\begin{description}
    \item[\textbf{1) Popularity.}] JavaScript is the main programming language for the Web. Due to web platforms' widespread presence these days, JavaScript has become one of the most popular programming languages in the world~\citep{JSpopular}. There are over 1.6 billion websites worldwide, and JavaScript is used on 95\% of them~\citep{JSInWorld}.
    %Virtually, all computing devices in use today run JavaScript, including iPhones, Android phones, Apple Mac OS, Microsoft Windows, Linux, smart TVs, and so on.
    According to GitHub, JavaScript is the programming language with the most GitHub projects~\citep{JSInGithub}. Also, we focused on popular JavaScript projects because they tend to be more prone to reuse, which may bring challenges to license usage.

    \item[\textbf{2) Availability.}] The JavaScript ecosystem is energetic and diverse, with more than 200k packaged readily available for mainstream use (on NPM). %Although many of these packages can be considered as trivial (with around 35 lines of code)~\citep{Abdalkareem:2017:FSE},
    The JavaScript community fosters a high reuse degree.
    About 10\% of the JavaScript packages available on NPM depend on at least one trivial package (with $\sim$ 35 lines of code)~\citep{Abdalkareem:2017:FSE}. On average, an ordinary JavaScript project published on NPM depends of other five packages~\citep{Meloca:2018:MSR}.

    \item[\textbf{3) Versatility}.] Other than web applications, JavaScript has also been frequently employed in other contexts such as mobile apps (iOS and Android)~\citep{JSMobileApps,Oliveira:2017:MSR}, database systems~\citep{dbJs}, operating systems~\citep{osJs},
    or even in embedded systems~\cite{JSIoT}. Moreover, JavaScript projects are frequently investigated in scientific studies~\citep{Meloca:2018:MSR,Campos:MSR:2019}.
\end{description}

Due to its popularity, availability, and versatility, JavaScript developers may have to deal with license issues more frequently than, say, developers in other traditional programming language communities.

We then downloaded the \totalProjects JavaScript projects locally.
Some notable open source projects, include:
\texttt{freeCodeCamp}\footnote{Available in: \url{https://github.com/freeCodeCamp/freeCodeCamp}},
%\texttt{bootstrap}\footnote{Available in: \url{https://github.com/twbs/bootstrap}},
\texttt{react}\footnote{Available in: \url{https://github.com/facebook/react}}, and
\texttt{angular.js}\footnote{Available in: \url{https://github.com/angular/angular.js}}.
Table~\ref{tab:qtdLinhasEProjetos} provides a quantitative
overview of the studied projects.

%\gnote{decrever um pouco mais dos projetos não-NPM}

\begin{table}[h]
	\centering
	\caption{A quantitative overview of the studied projects.}
	\label{tab:qtdLinhasEProjetos}
	\begin{tabular}{lrl}
    \hline Lines of code & \# Projects & Examples\\
    \hline
    %1 -- 100                 & 2     & react-devtools, lib-flexible\\
    1 -- 1,000             & 131   & localtunnel, parallel.js\\
    1,000 -- 10,000          & 558   & fetch, awesome-selfhosted\\
    10,000 -- 100,000        & 695   & flux, redux-thunk \\
    100,000 -- 1,000,000     & 168   & FreeCodeCamp, Angular.js  \\
    \hline
\end{tabular}
\end{table}

\subsubsection{Extracting License Information}
\label{analiseDados2}

To extract license related information, we resorted on a tool named ScanCode, created by NexB, a company that offers open source and third-party software scanning services. This tool implements an accurate license detection engine. According to \cite{dirkcomparison}, ScanCode has similar performance in detecting open source license, when compared to FOSSology~\citep{Gobeille:2008:MSR}, one of the longest living license scanning tool (over 83\% of all licenses).

%Instead of relying on regular expression or search patterns, it conducts a detailed comparison between any text data (that could contain license information) with a license text database and those found in the dataset analyzed.
To perform its analysis, ScanCode uses a large number of license texts and license detection ``rules'' that are compiled into a search engine.
When scanning the available text files, the text of the file is extracted and used to query the license search engine to find license matches. More concretely, ScanCode performs the following steps to detect open source licenses \footnote{Available in: \texttt{\url{https://scancode-toolkit.readthedocs.io/en/latest/explanations/overview.html\#how-does-scancode-work}}}:

\begin{enumerate}
    \item It collects an inventory of code files and sorts code files using file types;
    \item It extracts file content from any file using a general-purpose extractor;
    \item It extracts texts from binary files, if necessary;
    \item It uses a rule engine to detect warnings and open source license text;
    \item It uses a specialized parser to capture copyright claims;
    \item It identifies packages code and collects package metadata;
    \item It reports the results in JSON or CVS formats.
\end{enumerate}

We ran ScanCode over all \totalProjects JavaScript project repositories available in our dataset. The output of the tool is a CSV file with meta information, such as the number of licenses found, the name of the license, the files in which it found a license, etc., of the licenses found. For consistency purposes, we formatted the ScanCode output (i.e., the name of the licenses and their versioning) to the SPDX format\footnote{\url{https://spdx.org/licenses/}}.

\subsubsection{Double-checking ScanCode results}\label{sec:validacaoManual}

To validate the results that ScanCode reported,
we manually analyzed half (\totalProjectsAnaliseManual) of the results reported. We conducted this manual process as follows:

\begin{enumerate}
    \item For each CSV generated, we selected those files that were flagged containing license information;
    \item We accessed the online version of the project on GitHub and checked whether the flagged files indeed contain the license information reported by the tool;
    \item If the result was the same as reported in the tool, we marked an ``OK'' in a local spreadsheet;  we marked ``NOT OK'' otherwise.
\end{enumerate}

The first author conducted this process, and, in case of doubts, the last author was consulted. Examples of doubts included: when the author was not familiar with a license or when the same file had more than one license.

After this manual process, we observed that the tool could find correct license information for 95\% of the cases. The 5\% of the errors happened mostly when the license was briefly stated (i.e., the full license text was not provided; only the license acronym). For instance, in the file \texttt{README.md} of project \texttt{jquery.nicescroll}\footnote{\url{https://github.com/inuyaksa/jquery.nicescroll}}, the ScanCode tool reported the use of the GPL-2.0 license. Instead, we observed the use of the MIT license.

\subsection{Study 2: A Survey}\label{sec:survey}

Our survey had a broader goal. We first want to understand whether our respondent is aware of the multi-licensing issue, either at the project-level or at the file-level. While doing this analysis, we sought to cross-validate some of the findings observed in the mining study. Finally, we aimed to shed some additional light on understanding the reasons behind multi-licensing usage.

\subsubsection{Population}

Our target population comprises developers who made changes (i.e., commits) in license-related files of the studied JavaScript projects. By license-related files, we mean filenames that start with the ``license'' word (either with or without a text file extension).
After identifying these developers, we removed (1) those that are duplicated in our list (some developers have changed license files in different JavaScript projects) and (2) those with invalid email addresses. Our population then comprises \totalInvitedSurvey unique developers with valid email addresses that changed license-related files in the \totalProjects studied projects.

\subsubsection{Design}

Our survey was based on the recommendations of \cite{Kitchenham2008}. We followed the phases prescribed: planning, creating the questionnaire, defining the target audience, evaluating, conducting the survey, and analyzing the results. We set up the survey as an online questionnaire.
Before sending the actual survey, we conducted a pilot with five experienced researchers. We share with them our questionnaire, explained the purposes of our research, and explicitly asked them to fill the questionnaire and report back to us any problems and/or suggestions that they could have. We received feedback from three researchers that asked us to clarify some questions and improve the description of the questionnaire (available on the first page of the online form). We improved our questionnaire based on the feedback received.

After the modifications, the pilot responses were removed from the questionnaire, and we deployed the actual survey to our population of \totalInvitedSurvey developers (although 101 emails returned with error).
During a period of about 20 days, we obtained \totalFirstSurveyInvitation responses.
After about one month and a half, we sent a follow-up email inviting once again the participants to join in our study. Since the participants answered our questionnaire anonymously, we sent this follow up email to the entire population. After another 20 days from this second round of invitations, we received \totalSecondSurveyInvitation additional responses, totaling \totalSurvey responses in our survey.
Overall, we observed a $\sim$6\% of response rate. Our survey is available online at: \texttt{\url{https://forms.gle/PYbhWoREKK99VNPa9}}.

\subsubsection{Questions}
\label{subsec:questions}

Our survey had 11 questions (none was required, six were open). The questions covered in the survey are below:

\begin{itemize}
  \item[Q1.] How many public repositories do you own or have you contributed to? Choices: \{0, 1, 2--3, 4--5, ... , 11+.\}
  \item[Q2.] How many open source repositories did you choose a license for? Choices: \{0, 1, 2--3, 4--5, ... , 11+.\}
  \item[Q3.] How often do the projects you contributed to use more than one open source license? Choices: \{Always, Very often, Sometimes, Rarely, Never, I don't know.\}
  \item[Q4.] Are you aware of the implications of using more than one open source license? Choices: \{YES, NO\}
  \item[Q5.] If you answered YES to question \#4, could you please provide one example? \{Open\}
  \item[Q6.] How many times did you have to add more than one open source license to the projects you contributed to? \{0, 1, 2--3, 4--5, ... , 11+.\}
  \item[Q7.] If you answered anything other than zero in question \#6, WHY did you have to add more than one open source license? \{Open\}
  \item[Q8.] If you answered anything other than zero in question \#6, could you please describe (1) the license you added, (2) the license used before, and (3) why another license was important? \{Open\}
  \item[Q9.] Was there any situation in which you had to accept a contribution (aka pull request) that was licensed under a different license? Choices: \{YES, NO\}
  \item[Q10.] If you answered YES to question \#9, could you please elaborate it further with a concrete example? \{Open\}
  \item[Q11.] What are the files that you change when you need to add another license? \{Open\}
\end{itemize}

\subsubsection{Analysis}

We used two approaches for analyzing the answers to our survey. For the closed answers, we used basic descriptive statistics. For the open questions, we followed open-coding and axial-coding procedures~\citep{Strauss.Corbin_1998}.

\subsection{Replication Package}

%\fnote{Remover os links para replicação do estudo para garantir o blid review}
For replication purposes, all results generated in this study, including the scripts code used to perform the experiments, the data generated, and the survey questions and responses are available online on the Zenodo platform: \url{\zenodo}.

\section{Research Findings}

%After data collection, another Python script was created that analyzes the results and filters the results with more than one license listed. In total,
Before answering the research questions, we provide an overview of the data we collected.

\begin{figure}[t]
    \centering
    \includegraphics[scale=0.55,
    clip=true, trim= 0px 0px 0px 20px]
    {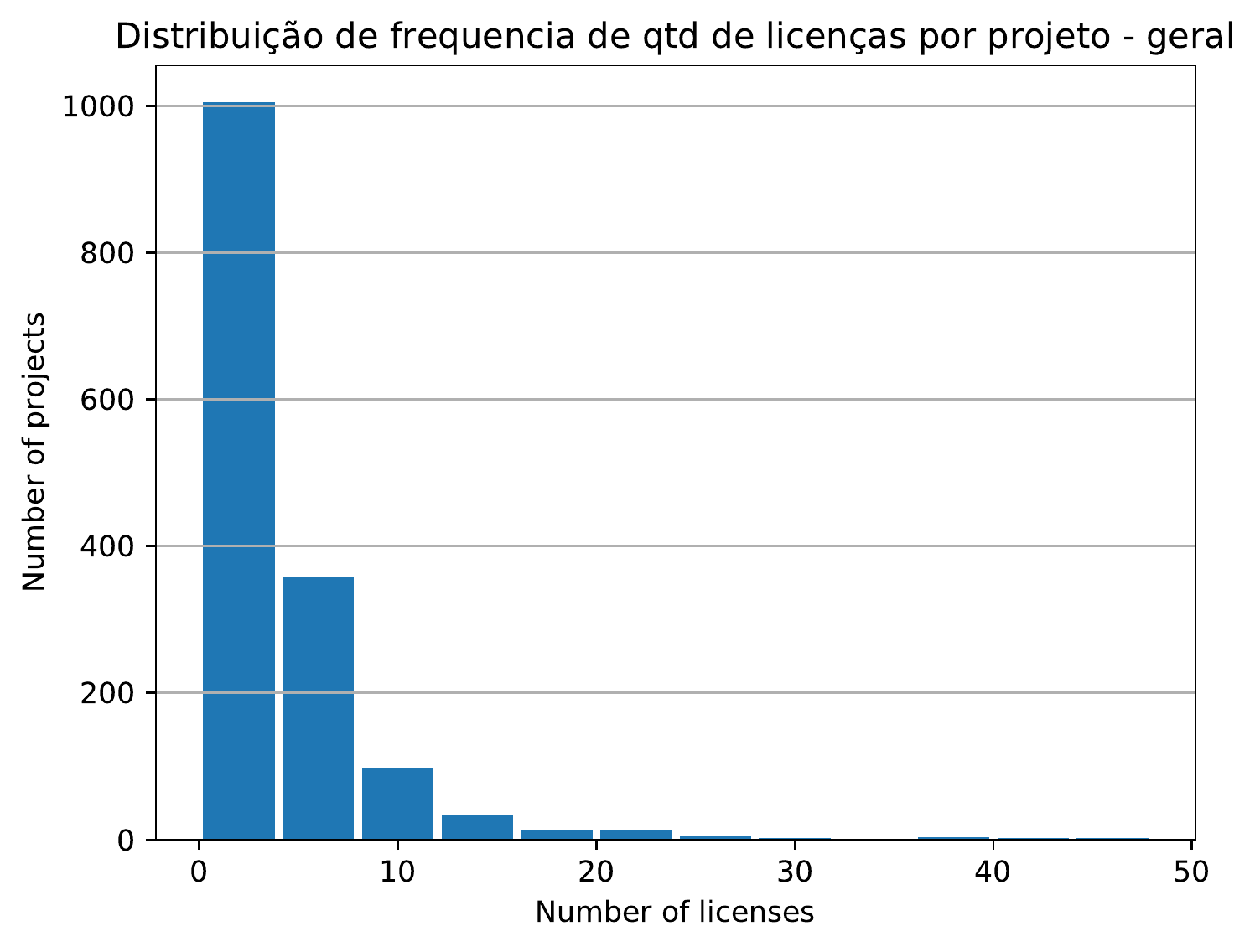}
    \caption{The distribution of the number of licenses found per project, considering all files in sub-directories. (Outliers have been removed so as not to damage the interpretation of the data).}
    \label{distribuicao-licencas-geral}
\end{figure}

After running ScanCode on the \totalProjects projects, we found a total of \totalLicenses instances of license usage; \totalUniqueLicenses individual licenses were found.
On average, there are 4.7 multiple licenses per studied project (median: 2, min: 0, max: 256, 3rd quartile: 5, standard deviation: 15.4).
There are a total of 16 projects for which we could not find any license declaration.
The distribution of the use of licenses is presented in Figure~\ref{distribuicao-licencas-geral}. Since this is a highly skewed distribution, the outliers were filtered out in this figure to ease data interpretation. However, the outliers were not removed from the raw data.

As mentioned, the distribution was highly influenced by a few outliers. Among them, we observed project \textsf{jxcore}\footnote{Available in: \url{https://github.com/jxcore/jxcore/}} (with 256 licenses), project
\textsf{node}\footnote{Available in: \url{https://github.com/nodejs/node}} (with 252 licenses),
and project \textsc{kibana}\footnote{Available in: \url{https://github.com/elastic/kibana/}} (with 240 licenses). %When studying the jxcore repository, we noticed that. One of the reasons noted is that there are many licenses that are inherited from project dependencies. \gnote{alguma razao pra ter todas essas licencas?}\jnote{Essa razão esta plausivel?}

Table~\ref{tab:resultadoLinhasDeCodigo} presents another perspective of the results, now organizing the studied projects in terms of lines of code.
In this table, it was noticed that small projects (up to 1,000 lines of code) use, on average, one license.
Similarly, on average, medium-sized projects (between 1,000 and 10,000 lines of code) use two licenses (558 projects). On the other hand, in large projects (with more than 100,000 lines of code), the average number of licenses per project is 19 (168 projects were found).

\begin{table}[t]
	\centering
	\caption{Description of license usage, organized in terms of lines of code (Projects with no license found were removed from this table).}
	\label{tab:resultadoLinhasDeCodigo}
	    \begin{tabular}{l c c c c c c}
        \hline Lines of Code & Projects & Mean & Median & Max & Min & St. Dev. \\
        \hline
        %1 -- 100 & 2 & 3,00 & 28 & 5 & 1 & 3\\
        1 -- 1,000 & 131 & 1.46 & 1 & 11 & 1 & 1\\
        1,000 -- 10,000 & 558 & 2.03 & 1 & 43 & 1 & 2 \\
        10,000 -- 100,000 & 695 & 4.12 & 3 & 42 & 1 & 4 \\
        100,000 -- 1,000,000 & 168 & 18.77 & 8 & 256 & 1 & 44 \\
        \hline
    \end{tabular}
\end{table}

\subsection{RQ1. How common are multi-licensed JavaScript projects?}

After running ScanCode on every source code file of each studied project, we observed that about 38\% of the projects (605 ones) have only one license, while most of them (\totalMultipleLicenses projects; 62\%) are multi-licensed.

However, it is well known that developers use different approaches for declaring open source licenses in their projects~\citep{Meloca:2018:MSR,GermanMSR}.
For example, while some developers declare their licenses as a comment in every source code file's header, other developers just cite the license used for the whole project within the \texttt{README} file. GitHub itself instructs developers to declare the project's license in a \texttt{LICENSE} file at the root dir of the project~\citep{LicensingARepository}.
Knowing that there are multiple approaches for declaring the use of open source licenses, we reran our experiments, but now focusing on multi-licensing at the project-level, that is, disregarding any other license that could be used in a subfolder.
%With this approach, we expect to identify the \emph{main licenses} declared in a project.

In this analysis, we observed a different scenario: only 30\% of the projects (497) are multi-licensed at the project-level; 1,055 of them use only one main license.
This suggests that developers do not declare all licenses used in the file-level also in the project-level, which may be essential to treat eventual license inconsistencies (more on this in  Section~\ref{sec:rq3}).
As a consequence, other developers interested in using such projects may inadvertently propagate license conflicts.

\begin{figure}[t]
    \centering
    \includegraphics[scale=0.55,
    clip=true, trim= 0px 0px 0px 20px]
    {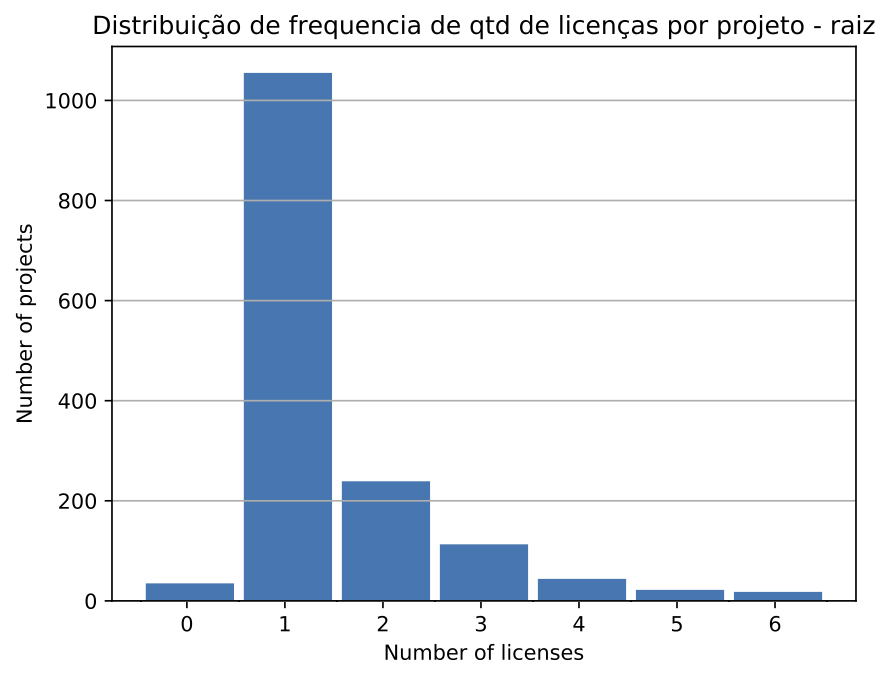}
    \caption{The distribution of multi-licensing at the project-level.}
    \label{distribuicao-licencas-raiz}
\end{figure}

Figure~\ref{distribuicao-licencas-raiz} reports the distribution of the licenses again but now considering project-level multi-licensing. As we can see, there are $\sim$1,100 projects declared with only one \emph{main} license. Moreover, there are $\sim$220 projects that adopt a dual license model. On the other hand, we found only 19 projects with more than seven \emph{main} licenses. This observation reinforces previous findings that many JavaScript projects do not declare all their licenses at the project-level.

\MyBox{\emph{\textbf{Finding \#1.}} When considering all source code files, we found a total of \totalMultipleLicenses multi-licensed JavaScript projects. However, when considering only multi-licensing at the project-level, we observed that 497 multi-licensed JavaScript projects.}

To better characterize multi-licensing at the file-level, we conducted a manual analysis of over half (\totalProjectsAnaliseManual) of the studied projects. In this analysis, we randomly sampled files in sub-directories of each project and investigated their license usage.
When conducting this analysis, we observed three main reasons for multi-licensing at the file-level:

\begin{description}
    \item[\textbf{Third-party libraries.}] Many of the licenses found at the file-level come from project's dependencies. For example, project \textsf{tablesaw}\footnote{Available at: \url{https://github.com/filamentgroup/tablesaw/}}, which is a library that implements plugins for responsive table creation, declares the use of the MIT license in its \texttt{LICENSE} file and its \texttt{package.json} file, but also makes use of the Apache-2.0 license in some of its sub-directories. This happened because the source code of external libraries were added to the codebase. For instance, this \textsf{tablesaw} project has the library \textsf{qunit.js} included as a dependency. This particular \textsf{qunit.js} library, on the other hand, is licensed under MIT and Apache-2.0 licenses.

    \item[\textbf{Documentation.}] Some projects employ documentation licenses, such as the Creative Commons family of licenses. For instance, we observed that the \textsf{angular.js} project, a framework maintained by Google that assists the creation of single-page applications, declares the use of the MIT license in its \texttt{LICENSE} file and in the \texttt{package.json} file, but also uses a Creative Commons license in some of its sub-directories, to license documentation files.

    \item[\textbf{Examples.}] The use of a license in source code ``how to'' examples. For instance, the \textsf{rocket.chat}\footnote{Available at: \url{https://github.com/RocketChat/Rocket.Chat}}, an open source team communication platform, declares the use of a MIT license in the \texttt{LICENSE} file and in the \texttt{package.json} file. However, it also uses Apache-2.0 to license sample files.
\end{description}

The use of third-party libraries in sub-directories seems to be a strong indicator for the presence of multi-licensing at the file-level. This is particularly relevant to the JavaScript ecosystem since JavaScript projects could be \emph{minified}, that is, the process of removing unnecessary characters (e.g., white space characters, newline characters, comments, etc.) from source code without changing its behavior. The purpose of minification is to reduce the size of the source code. Since JavaScript projects are predominantly executed in browsers, reducing its size also means improving user experience (i.e., a webpage would load faster).

To shed additional light on this matter, we selected three popular libraries in our dataset (\texttt{bootstrap}, \texttt{jquery}, and \texttt{angular}) and investigated how many times they appear as minified libraries in other JavaScript projects in our dataset. We observed that there are 64, 87, and 39 JavaScript projects that use the minified version of the \texttt{bootstrap}, \texttt{jquery}, and \texttt{angular} libraries, respectively.
%cat allResults.csv  | grep "angular.*min.js" | awk -F/ '{print $2}' | sort | uniq | wc -l

When searching for source code files with the ``.min.js'' suffix, we found 481 JavaScript projects (out of the \totalProjects studied ones) that employ at least one minified JavaScript library. Since each one of the minified version of these projects declare their license in the header of their files, the JavaScript projects that use these libraries also incorporate their license.

\MyBox{\emph{\textbf{Finding \#2.}} Many JavaScript multi-licensing at the file-level happens due to the reuse minified version of other JavaScript libraries, and these libraries are released under their own licenses.}

\subsection{RQ2. What are the pairs of licenses most used together?}

In this research question, we group the results regarding pairs of software licenses and pairs of non-software licenses.
We combined the ten pairs of licenses that appear more frequently to shed some light on this matter.

\subsubsection{Pairs of software licenses}

Table~\ref{tab:pairs_of_license} shows the results of the ten most common pairs of software licenses. To build this table, we first selected the source code files under different licenses per project; we then created pairs with the most common ones and finally sorted them out.
It is important to note that, in this analysis, we are not considering multi-licensing at the project-level (which would greatly reduce our data, since only 30\% of the projects are multi-licensed at the project-level); we focused on the file-level, instead.

\begin{table}[h]
	\centering
	\caption{The list of the pair of licenses that are mostly used together. }
	\label{tab:pairs_of_license}
	    \begin{tabular}{l c l}
        \hline Licenses Used & \# Occurrences & Examples\\
        \hline
        BSD-3 \& MIT & 404 & \textsf{annyang}\\
        Apache-2.0 \& MIT & 340 & \textsf{elasticsearch-hq}\\
        Apache-2.0 \& BSD-3 & 221 & \textsf{nvd3}\\
        GPL-2.0 \& MIT & 194 & \textsf{vibrant.js}\\
        GPL-1.0-plus \& MIT & 188 & \textsf{qunit.js}\\
        Apache-2.0 \& GPL-2.0 & 120 & \textsf{vuejs}\\
        Apache-2.0 \& GPL-1.0-plus & 117 & \textsf{cat} \\
        GPL-1.0-plus \& GPL-2.0 & 113 & \textsf{gateone} \\
        \hline
    \end{tabular}
\end{table}

% pra produzir a tabela acima
% sort -k3 -n -r file_output_parser.py | grep -v -E 'public|ofl'  | head -n10

As one could see, the pair ``BSD-3 \& MIT'' is the most frequent used one, appearing 404 times. This means that there are 404 projects that have at least one source code file licensed under MIT and another source code file licensed under BSD-3. The next most common pairs are ``Apache-2.0 \& MIT'' and ``Apache-2.0 \& BSD-3''. Interestingly, these three licenses (MIT, Apache-2.0, and BSD-3) are essential ones because they allow derivative works with very few restrictions (even with different terms or licenses), as long as the necessary credit is given to the original work~\citep{BSDeMIT}.
In particular, the Apache-2.0 license is a derivative of the BSD license and is also similar to the MIT license~\citep{Kechagia:2010}. This may indicate why these licenses are frequently employed together.

It is also interesting to note the GPL (version 1 and 2) license and the MIT license as part of the most common pairs of licenses. One vivid reader would wonder if the use of GPL, for instance, version 2, is allowed with the use of the MIT license. To answer this question, we first have to figure out the license defined at the project-level. We leave the analysis to RQ3.

%(along with the MIT license). This particular license is not a software license and is not approved by either OSI or FSF, and it is not even listed in SPDX, a consortium of non-profit and profit organizations that attempt to standardize licensing information across parties\footnote{https://spdx.org/licenses/}.

\MyBox{\emph{\textbf{Finding \#3.}} BSD-3 license and the MIT license compose the most common pair of licenses in the studied projects, followed by the pair Apache-2.0 and MIT pair, and the Apache-2.0 and BSD-3 pair. Even though all these three licenses are permissive, they have different compatibility levels to copyleft licenses such as GPL.}% This is one of the reasons to release a project by using multiple licenses.}

\subsubsection{Pair of non-software licenses}

%As noted earlier, some of the licenses found in sub-directories were not specific software licenses (such as the Creative Commons license). To better understand the existence and use of non-software licenses, we summarized the licenses used in terms of software licenses and non-software licenses (documentation licenses, patent, etc.).

To make sure that a license is a software license or a non-software license, we relied on specialized online mediums such as the OSI website\footnote{\url{https://opensource.org/licenses}}, the DejaCode website\footnote{\url{https://enterprise.dejacode.com/licenses/}}, and ChooseALicense website\footnote{\url{https://choosealicense.com/non-software/}}, that provide a comprehensive description of these licenses, as well as search engines.
When considering only licenses recognized by SPDX, we found a total of \totalUniqueNonSoftwareLicenses unique non-software licenses used in our studied dataset of JavaScript projects.

%cat all_results_w_spdx.csv |  grep -v -E 'mit|apache|bsd|gpl|mpl|epl|isc|w3c|openssl*|softsurfer|ekioh|svndiff|twisted-snmp|intel|wtfpl*|zlib|boost-1.0|dco-1.1|info-zip-2007-03|python|json-pd|efl-2.0|x11*|curl|nrl-permission|afl-2.1|boost-1.0|lppl*|bsla|apafml|adapt*|zpl*|ypl*|xfree86*|uoi-ncsa|upl-1.0|vsl-1.0|wordnet|regexp|reportbug|robert-hubley|rsa-1990|rsa-md5|sax-pd|scea-1.0|sgi-freeb-1.1|sgi-freeb-2.0|sgi-fslb-1.0|sparky|st-mcd-2.0|sugarcrm-1.1.3|sunpro|sun-sissl-*|sybase|taligent-jdk|tatu-ylonen|android-sdk-preview-2015|amazon-sl|bloomberg-blpapi|boost-original|bpmn-io|arphic-public|bitstream|elastic-license-2018|jetty|jquery-pd|json-js-pd|rsa-1990|taligent-jdk' | awk -F, '{print $2}'| sort | uniq | wc -l

These \totalUniqueNonSoftwareLicenses have been used to license a total of 4,544 different files (out of the \totalFiles files parsed; 0.3\%).
Among the most common non-software licenses found, we highlight the SIL Open Font License 1.1 (OFL-1.1 for short), which is by far the most used non-software license, licensing a total of 2,579 (56\%) files. The Creative Commons family of licenses comes next: Creative Commons -- Public Domain (cc-pd) is used in 287 files, Creative Commons Attribution version 4 (cc-by-4) is used in 245 files, and Creative Commons Attribution version 3 (CC-BY-3) is used in 236 files.
In the project-level, we found that 275 (out of the \totalProjects; 17\%) JavaScript projects employ at least one non-software license. All these projects also use at least one software license, meaning that we found no JavaScript project that only employs non-software licenses.

Project \textsf{mathjax}\footnote{Available at \url{https://github.com/mathjax/mathjax}} is the one with the highest number of files licensed under non-software licenses: a total of 765 files. When looking closer at this particular project, we noticed a JavaScript engine for LaTeX (and other typesetting systems) that works in the browsers. \textsf{mathjax} included several font files; all of them licensed under OFL-1.1.

\MyBox{\emph{\textbf{Finding \#4.}} Only 0.3\% of the files are licensed under a non-software license. About 17\% of the projects employ at least one non-software license (at the project-level). The use of the OFL license represents the majority of the non-software license usage.}

\begin{table}[H]
	\centering
	\caption{A list of the 10 most common pairs of licenses that include at least one non-software license.}
	\label{tab:pairs_of_non_sw_license}
	    \begin{tabular}{l r l}
        \hline Licenses Used & \# & Examples\\
        \hline
        %mit & public-domain & 129 \\ SPDX nao reconhece
        MIT \& OFL-1.1 & 124 & \textsf{elasticsearch-head} \\
        Apache-2.0 \& OFL-1.1 & 94 & couchdb\\
        %BSD-3 & public-domain & 88 \\  SPDX nao reocnhece
        BSD-3 \& OFL-1.1 & 84 & kibana\\
        %apache-2.0 & public-domain & 77 \\
        %free-unknown & mit & 73 \\
        %generic-cla & mit & 65 \\
        %mit & unknown-license-reference & 61 \textsf{\\
        CC-BY-4.0 \& MIT & 59 & \textsf{etcher}\\
        CC-BY-3.0 \& MIT & 58 & \textsf{bootstrap}\\ %\footnote{\url{https://github.com/jasny/bootstrap/}}\\
        %gpl-1.0-plus & public-domain & 57 \\
        %gpl-2.0 & public-domain & 55 \\
        %BSD-3 & facebook-patent-rights-2 & 54 \\
        %apache-2.0 & free-unknown & 54 \\
        %BSD-3 & free-unknown & 52 \\
        %facebook-patent-rights-2 & mit & 49 \\
        CC-BY-SA-3.0 \& MIT & 48 & bootstrap-table\\
        GPL-2.0 \& OFL-1.1 & 47 & elasticsearch-HQ\\
        %apache-2.0 & generic-cla & 47 \\
        CC-BY-SA-4.0  \& GPL-1.0-plus & 14 & hyper\\
        CC-BY-SA-3.0  \& CC-BY-SA-4.0 & 14 & anti-adblock-killer\\
        CC-BY-4.0  \& LGPL-2.1-plus & 14 & sharp\\
        \hline
    \end{tabular}
\end{table}

Table~\ref{tab:pairs_of_non_sw_license} shows the most common pair of licenses that include at least one non-software license. In this table, we only included non-software licenses that are recognized by SPDX. As one could observe, the most common pairs are  ``MIT \& OFL-1.1'', followed by ``Apache-2.0 \& OFL-1.1'', and ``BSD-3 \& OFL-1.1''. The Creative Commons family of licenses (e.g., CC-BY-SA, CC-BY-3.0, CC-BY-4.0, etc.) is also frequently employed, mostly with the MIT license.
This table is in sharp agreement with Table~\ref{tab:pairs_of_license}, which shows that the combination of the MIT, Apache-2.0, and the BSD-3 licenses are among the most common pair of (software and non-software) licenses. Here we also observed the same three licenses again, but now pairing with the OFL-1.1 license.

Interestingly, when reading the definition of the OFL-1.1 license\footnote{Available at \url{https://scripts.sil.org/cms/scripts/page.php?item\_id=OFL10\_web}}, we noticed that this license has the goal of ``stimulating a worldwide development of cooperative font projects, [..] providing an open framework in which fonts may be shared and improved in partnership with others.'', it could also license file types, including: (1) font files, (2) data files, (3) source code, (4) build scripts, and (5) documentation. Indeed, this license was approved by OSI as a software license\footnote{\url{https://opensource.org/licenses/OFL-1.1}}.
Digging further to understand the OFL-1.1 license usage, we investigate the file extensions of these source files licensed under it. Among the 2,834 files licensed with OFL-1.1, we found that 1,119 of them are \texttt{.ttf} files (a file extension for a font file developed by Apple), 711 are \texttt{.eot} files (for font files developed by Microsoft), 611 are \texttt{.otf} file (for font files developed by Adobe and Microsoft). Only 79 of them are \texttt{.css} files and 63 of them are \texttt{.js} files.

\MyBox{\emph{\textbf{Finding \#5.}} Among the pair of (software and non-software) licenses, MIT, Apache-2.0, and BSD-3 are often employed along with the OFL license. The Creative Commons family of licenses is also very frequently used.}

%- Tem algum projeto licenciado apenas com licenças de documentação?
%- Será que licenças de documentação são usadas em arquivos de codigo?

% COMPATIBILIDADE DE LICENÇAS ---------------------------------------------------

\subsubsection{More than a pair of licenses?}

We performed another round of experiments to cover different cardinalities, such as triples instead of pairs. However, while conducting this experiment, we noticed that triples are not fairly common. The most common 3-license sets we found has nine occurrences.
Some examples include: BSD-3 \& CPL-1.0 \& GPL-2.0-plus (9 occurrences), GPL-3.0-plus \& LGPL-3.0 \& MIT (9 occurrences), and Apache-2.0 \& CC-BY-3.0 \& LGPL-3.0 (9 occurrences).
This means that there are only nine projects that have three different source code files licensed under BSD-3, CPL-1.0, and GPL-2.0-plus.

\subsection{RQ3. How common are license incompatibility issues in multi-licensed open source projects?}\label{sec:rq3}

To investigate license compatibility issues, in this research question, we used the notion of compatibility introduced by~\cite{Kapitsaki:JSS:2017}. \cite{Kapitsaki:JSS:2017} states that ``license compatibility stems from a less restrictive to a more restrictive license. For instance, although the MIT and the BSD licenses seem equivalent, software the contains both licenses should be licensed under the more restrictive license. Using the less restrictive license would cause violations due to the lack of conformance to obligations present in the more restrictive license. ``More concretely, License A is considered
``one-way compatible'' with license B, if software that contains components from both licenses can be licensed under license B. The term ``one-way'' is used because although license A can be compatible with B, the opposite does not always hold.

Since license compatibility issues should consider different variables (e.g., whether a component with different license is included dynamically or statically in derivative works), we opted for a more conservative approach: we used a subset of the graph of license compliance provided by \citep{Kapitsaki:JSS:2017}. Table~\ref{tab:incompatible_licenses} describes incompatibility examples. We also explored additional combinations of licenses, but we only report those that we found any occurrences in the studied JavaScript projects.
This table should be read as: there is a license incompatibility if an open source software is licensed under MIT (in the project-level) and has any component (in the file-level) that is licensed under BSD-3, or Apache-2.0, or  MPL-1.1, etc.

%To build this table, we first searched for projects in which the \emph{main} license is in the set present at the "Licensed under" column. We then searched for source code files within these projects licensed under any of the licenses presented at the "Incompatible with" column.

\begin{table}[h]
	\centering
	\caption{License Incompatibility issues. The symbol $\lor$ refers to the `OR' logical operator.}
	\label{tab:incompatible_licenses}
    \begin{tabular}{l p{8cm} r}
    \toprule
    Licensed under & Incompatible with & \#\\
    \midrule
    MIT & BSD-3 $\lor$ Apache-2.0 $\lor$  MPL-1.1 $\lor$ MPL-2.0 $\lor$ LGPL-2.1 $\lor$ LGPL-2.1+ $\lor$ LGPL-3.0 $\lor$ GPL-2.0 $\lor$ GPL-2.0+ $\lor$ GPL-3.0 $\lor$ AGPL-3.0  & 309 \\

    BSD-3 & Apache-2.0 $\lor$ MPL-1.1 $\lor$ MPL-2.0 $\lor$ LGPL-2.1 $\lor$ LGPL-2.1+ $\lor$ LGPL-3.0 $\lor$ GPL-2.0 $\lor$ GPL-2.0+ $\lor$ GPL-3.0 $\lor$ AGPL-3.0 & 31 \\

    Apache-2.0 & MPL-1.1 $\lor$ MPL-2.0 $\lor$ LGPL-2.1 $\lor$ LGPL-2.1+ $\lor$ LGPL-3.0 $\lor$ GPL-2.0 $\lor$ GPL-2.0+ $\lor$ GPL-3.0 $\lor$ AGPL-3.0 & 39 \\

    %MPL-1.1 & LGPL-2.1 $\lor$ LGPL-2.1+ $\lor$ LGPL-3.0 $\lor$ GPL-2.0 $\lor$ GPL-2.0+ $\lor$ GPL-3.0 $\lor$ AGPL-3.0 & 0 \\

    MPL-2.0 & LGPL-2.1 $\lor$ LGPL-3.0 $\lor$ GPL-2.0 $\lor$ GPL-2.0+ $\lor$ GPL-3.0 $\lor$ AGPL-3.0 & 4 \\

    %LGPL-3.0 & GPL-2.0 $\lor$ GPL-2.0+ $\lor$ GPL-3.0 $\lor$ AGPL-3.0 & 0 \\
    GPL-3.0 & GPL-2.0 & 8\\
    GPL-2.0 & GPL-3.0 & 2\\
    %AGPL-3.0 & GPL-3.0 & 0 \\
    \bottomrule
    \end{tabular}
\end{table}

We found license inconsistencies in 372 projects (out of the \totalMultipleLicenses multi-licensed projects; 39\%).
However, most of the license inconsistencies happen with projects licensed under MIT license (309 occurrences). Overall there are 636 JavaScript projects licensed under MIT. When digging a bit further, we noticed that 288 (out of the 309 inconsistencies with MIT projects; 93\%) are related to the use of source code files licensed under BSD-3 or Apache-2.0 licenses. This license inconsistency happens because MIT is less restrictive than BSD-3 and Apache-2.0~\citep{laurent2004understanding}.
Moreover, according to \cite{Kapitsaki:JSS:2017}, any software that contains multiple licenses (in the file-level) should be license under the more restrictive one (in the project-level).
To give a more general perspective, we explored copyleft licenses (e.g., GPL-2.0, GPL-2.0+ or GPL-3.0) in the file-level within JavaScript projects licensed under a permissive (e.g., MIT, BSD-3, or Apache-2.0) in the project-level. We found that 154 JavaScript projects with this license incompatibility.

On the other hand, we observed a small number of license inconsistencies when considering JavaScript projects licensed under copyleft licenses (e.g., GPL-2.0, GPL-3.0, and AGPL-3.0). This finding is mainly due to the also small number of projects licensed under these licenses. Overall there are 52 JavaScript projects licensed under either GPL-2.0 (15 projects), GPL-3.0 (27 projects), or AGPL-3.0 (10 projects).

Finally, one reader might found strange that GPL-2.0 is not compatible with GPL-3.0 (and the other way around). According to GNU Project\footnote{\url{https://www.gnu.org/licenses/license-list.html\#GNUGPLv3}}, GPL-3.0 and GPL-2.0 are, indeed, incompatible. To avoid this concern, developers can apply the sentence ``or later'' to some version of the GPL, like ``GPL-2.0 or later'', or simply ``GPL-2.0-plus'' or even ``GPL-2.0+''.

\MyBox{\emph{\textbf{Finding \#6.}} We found license inconsistencies in 39\% of the multi-licensed JavaScript projects. The most common inconsistency happens when a project licensed under MIT at the project-level use restrictive licenses in the file-level. There are very few license incompatibility issues with the GPL family of licenses; mainly because these licenses are not in widespread use in our dataset.}

% LOCALIZAÇÃO DAS LICENÇAS ---------------------------------------------------
\subsection{RQ4. Are multi-licensed projects warning their potential clients on the use of multiple licenses?}

To understand license declaration at the project-level, we initially selected three well-known and widely used files for this purpose: the \texttt{LICENSE} file, the \texttt{README} file, and the \texttt{package.json} file.

The \texttt{LICENSE} file, as its name suggests, is the most common file for declaring software licenses. Its use is extremely widespread, and development platforms such as GitHub use this file to automatically infer the licenses employed. The \texttt{README} file, on the other hand, provides an overview of the project, usually for new developers who want to get used and/or contribute to the project. This file is often used to declare the licenses of a project, although it is not necessarily required. However, maintainers believe this is an important practice because external visitors often access this file. Finally, the \texttt{package.json} file is the gateway to NPM, the Node package manager system. In addition to the license definition, this file states information such as the project description, the project version, the repository URL, the production and development dependencies, among other things.
Suppose a license is declared after deploying the package to the NPM infrastructure. In that case, users browsing the NPM website will be able to straightforwardly identify the license used directly on the NPM website without browsing the source code or accessing other websites (such as GitHub). In any case, it is the project maintainer's responsibility to add license information to these files.

\begin{figure}[tp]
    \centering
    \includegraphics[scale=0.55,
    clip=true, trim= 0px 0px 0px 20px]{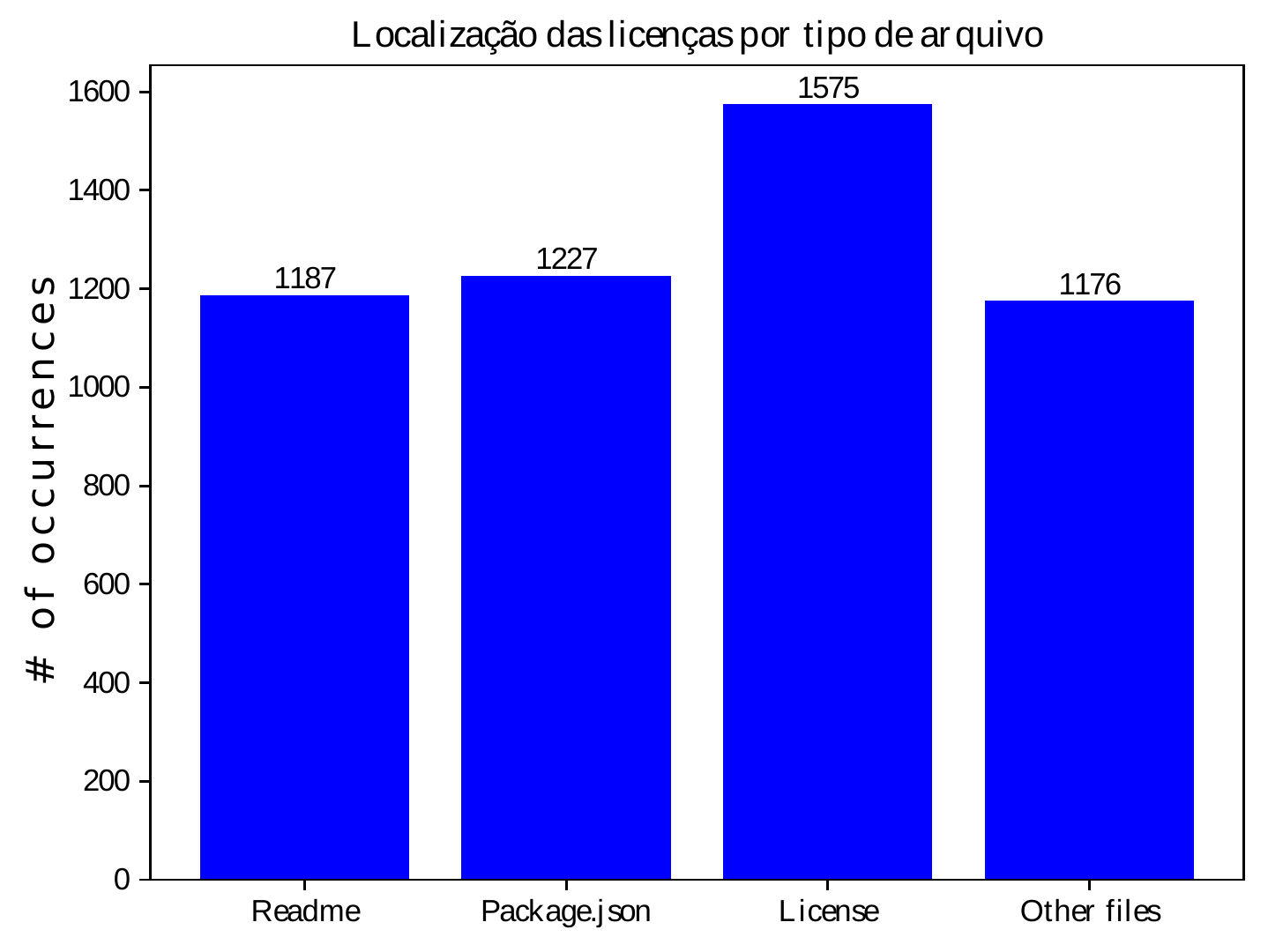}
    \caption{Files in which licenses are declared.}
	\label{localizacao-licencas-raiz}
\end{figure}

Figure~\ref{localizacao-licencas-raiz} shows the distribution of licenses among the mentioned files. The ``other files'' bar, means any other file other than the mentioned ones. In this particular experiment, however, focused on multi-licensing at the project-level.

The first remarkable observation from Figure~\ref{localizacao-licencas-raiz} is that there is a disagreement on how developers declare their license usage. As one can observe, most of the license declarations were found in the \texttt{LICENSE} files. One avid reader could also note that the number of \texttt{LICENSE} files is greater than the overall number of studied JavaScript projects (1,575 \texttt{LICENSE} files vs. \totalMultipleLicenses projects with more than one license).
This happened because some of the projects that employ multiple licenses list each one of these licenses in an individual \texttt{LICENSE} file. An example is the Annotator project\footnote{Available at \url{https://github.com/openannotation/annotator}}, a library for building web browser annotation apps. This particular project declares a MIT license in a file named ``\texttt{LICENSE-MIT}
'' and a GPL license in another file called ``\texttt{LICENSE-GPL}''.
Moreover, other known files, such as the \texttt{COPYING} file are not used frequently for license declaration: we observed only 15 instances of the \texttt{COPYING} file being used for this purpose.

Finally, to better understand the ``Other files'' bar, we manually inspected these instances. While doing so, we observed an interesting pattern: some JavaScript projects declare their license usage using a file named with the name of the license. For instance, project \textsf{jquery-ujs}\footnote{\url{https://github.com/rails/jquery-ujs/}} created a file named \texttt{MIT-LICENSE}, whereas project \textsf{kibana}\footnote{\url{https://github.com/elastic/kibana/}} declare its license usage with a file named \texttt{APACHE-LICENSE-2.0.txt}. Other projects use the same pattern with other abbreviated licenses.

\MyBox{\emph{\textbf{Finding \#7.}} As one could expect, the \texttt{LICENSE} file is the most common used for declaring multi-licensing. However, these definitions are not accurately distributed among other important files.}

\subsection{RQ5. Are developers aware of multi-licensing issues in their projects?}\label{sec:rq5}

In this section, we report the analysis of the \totalSurvey answers to our survey.

We observed that 77\% of our 83 respondents contributed to more than 11 projects regarding our population. Beyond that, 95\% of them were somehow involved in choosing their project's license(s); we highlighted that 77\% of our respondents were engaged in selecting licenses for more than one project that they contribute to. Surprisingly, $\sim$63\% claimed that they are not aware of the implications of using more than one license in their projects (either at the project-level or at the file-level). Moreover, although 28 of 83 respondents provided explanations for their multi-licensing awareness, we also found controversial answers. One respondent mentioned that
``everything you contribute is dual licensed automatically and users may chose any of the dual-licenses''. Another respondent mentioned that ``it's an `OR' condition; so anything that's permitted by *any* of the licenses, is permitted.''. These answers showed that the use of multiple licenses might confuse software developers.

%, "Contributing to a project with a dual open source and proprietary license requires contributors to acknowledge their contributions will be used for proprietary software" and, "license compatibility is one issue: just because two licenses are free software licenses, it doesn't mean they are compatible.".

%This fact was pointed in the question Q5 of the survey when a developer told "They need to be compatible. For instance MIT and BSD are compatible with the Apache license but GPL is not" or in the following answer:

%\begin{quote}
%\emph{ "It is possible to dual license a project with licenses that have different permissiveness e.g. GPL + Apache-2.0. This could allow folks to use the software with the license they find more agreeable."}
%\end{quote}

As for the reasons for multi-licensing at the project-level, we obtained 13 valid answers to questions seven and eight. In two of these answers, developers did not provide much detail, such as ``multiple cases''. For the remaining answers, four were related to \textbf{``solving the license compatibility''}. For example, one of the developers claimed that ``Wordpress was looking to use my library and the Apache 2.0 license is incompatible with GPL v2
''. The solution adopted was to include a new license in the project.

Four developers argued that their intention was to \textbf{``add flexibility to the existing license''}. In one of the cases, the library adopt the ``OpenSSL exception clause'' license into a GPL project. In another case, the ``existing license was too limiting, and the rest of that company's licenses had already been updated to be less limiting''. The most common license combinations presented in Table~\ref{tab:pairs_of_license} (except the last one) allow the use of the library in both copyleft and permissive software. This happens when the pair has at least one permissive license such as BSD or MIT, or a permissive license and a copyleft (GPL) license.
The last reason raised in the survey was related to \textbf{``legal third-party adoption''}. Two developers mentioned that a license was added to the project to satisfy the Apache Software Foundation requirements. Another respondent noted that ``as the source release contained third-party code under a compatible license, this is distinct from dual licensing.''

\MyBox{\emph{\textbf{Finding \#8 }} We found three reasons on why developers adopt multiple licenses at the project-level, including: the importance to take care of license compatibility issues, legal third-party library adoption, and making the project less restrictive by adding another license.}

Regarding copyleft/GPL compatibility issues, in the project-level, it is possible to adopt a GPL license along with a permissive license. Some developers shared this approach that answered Q5, for example, ``some people need projects licensed in the GPL family, and some people need projects licensed in NOT the GPL family. For those projects that don't care either way, sometimes they are dual licensed'', or ``this code is dual-licensed under the MIT and GPL-3.0 licenses. This means that you, as user, may choose one of these licenses to abide by. If complying with the GPL is problematic for you, you can choose the more liberal MIT license''.

When we asked the respondents to point out cases when they accepted contributions that may add additional licenses in the file-level, five respondents mentioned that they have accepted pull requests in which: (i) ``some contributions \textbf{introduce new dependencies} that sometimes have different licenses, so there is a judgment call to me made whether to include the contribution''; (ii) that her
``\textbf{copy and pasted} code with different but very similar license'', (iii) that the respondent had seen ``contributors send PRs that included GPL licensed (or similar copyleft) dependencies, which were merged to an MIT project. Once the \textbf{issue [with the licenses]} was noticed, the dependencies were removed and PRs reverted.'' ; and two respondents mentioned that some ``[...] contribution(s) arrived \textbf{without any license}''.

\MyBox{\emph{\textbf{Finding \#9 }} Some project maintainers accept contributions that involve multi-licenses, although they acknowledge that extra care should be placed in these situations.}

Finally, to provide additional analysis, we conducted a manual investigation over a random sample of 195 commits ((confidence level of 95\% with a ±5\% error)) that modified license-related files. We perform this analysis to understand if the commit message or any other changes could provide us any hint on the decision of using multiple licenses.

From this inspection, we identified that the majority of the commits either created the \texttt{LICENSE} file (73 commits), updated the license date (65 commits), or changed the author's copyright (18 commits). We found 14 re-licensing occurrences (e.g., from BSD-2 to GPL, MIT to Apache 2.0, BSD-3 to MIT, etc.). We also discovered 18 commits in which developers made minor changes in the license text, for example moving lines, or adding extra information. Unfortunately, only seven commits showed some relation to multi-licensing. All of them were related to \textbf{
``legal third-party library adoption''} category aforementioned.
%, which happens when developers make clear that the libraries used in file-level can have different licenses when compared to licenses in the project-level.

\MyBox{\emph{\textbf{Finding \#10 }} Through the manual analysis over the commits in license-related files, we observed that the majority of the changes are not related to multi-licensing.}

\section{Discussion}

In this section, we provide an additional discussion on the data presented in the previous sections.

%There are several ways for developers could violate license usages, such as reusing code licensed under GPL Version 2 license or higher in a project that is licensed under MIT license. We explore two kinds of violation: (1) the use of unknown licenses, that is, licenses that are not recognized by the Software Package Data Exchange (SPDX). We decided to focus on SPDX because it has a workgroup hosted by The Linux Foundation that catalogs the most comprehensive list of open source licenses available to the best of our knowledge. (2) the use of at lease one permissive license and one reciprocal license in the same project.\fnote{Então, na verdade só haverá violação se o projeto for permissivo e algum código em reuso for recíproco. O oposto não configura uma violação.}\gnote{descrever algo aqui..}
%Particularly important for this work, we are interested in using licenses that have not been formally approved by an entity (Section ~\ref{sec:spdx}), as well as the unrestricted use of permissive and restrictive licenses (Section ~\ref{sec:comp}). In this topic the analysis was performed in projects that had more than two licenses.

\vspace{0.2cm}
\noindent
\textbf{Many software licenses are not recognized by SPDX.}
Among the \totalUniqueLicenses distinct licenses found, it was observed that SPDX did not recognize about 35\% (146) of them.
Among the most frequently used licenses not recognized by SPDX (of a total of \totalLicenses license occurrences) there are the Facebook Software License (with 10,329 occurrences), the Unicode Inc License Agreement (with 994 occurrences), and Google licenses Patent License for WebM (with 212 occurrences).
This occurs because the primary concern of SPDX is about open source and free software licenses specifically, not about software licenses in general. Despite most of those licenses allowing the use and redistribution of the licensed project, they are not classified as a free software license because other freedoms are not guaranteed by them, such as the freedom to redistribute the code.

%\vspace{0.2cm}
%\noindent
%\textbf{Dual-license: free and proprietary software.} The availability of a software by a combination of free software license (mainly some copyleft licenses) and a proprietary license was a commercial model experimented by several companies in the past, such as Trolltech, Sun, MySQL and others. This theme was also explored by researchers in other research works~\citep{valimaki2003dual, koski2005oss}. Despite this topic was not covered by our analysis, it is important to point out that this kind of availability was mentioned by some participants of our survey. For question Q5, for example, a participant told that "[we can] offer a free, copyleft version and a non-free commercial version of the same software". In the same sense, other participant told  that "a secondary license allows for withholding some value under particular conditions, for example GPL disallows commercial use and requires share alike rights, but a secondary proprietary license can release those rights under a contract.". Those answers are aligned with some companies using this same commercial strategy. For example, the Qt framework developed by Digia is available in both LGPL and commercial licenses\footnote{The licensing model of Qt can be found in \url{https://www.qt.io/licensing/}}, reminding us that this reason for dual licensing is a recurrent theme in free software and open source community.

\vspace{0.2cm}
\noindent
\textbf{JavaScript development: package managers.} In the search for productivity, package managers brought developers new functionalities concerning ease of use and reuse of preexisting software code. Regarding JavaScript, this new development context has introduced new practices. NPM, considered one of the most popular JavaScript package managers, have set a development practice that might seem uncommon to other languages. When including libraries onto a project, the source code is imported into the project's source code. In general, this behavior is to assure that a specific version of the library is strongly tied to the project, avoiding issues when library maintainers change or update their code. Considering that each library may choose its own open source license accordingly, the importing action brings the source code license to the project using it. Thus, a small JavaScript project using NPM has a high number of different software licenses (including open source and non-open source licenses) very quickly. Yet, this methodology does not seem to be completely understood by developers, since our findings show that most projects we analyzed use a single license for the project, ignoring that licenses of imported libraries might conflict with their project's license. An example is the most common pair of software licenses used together: BSD-3 \& MIT, which has 404 occurrences, according to Table~\ref{tab:pairs_of_license}. Note that projects licensed under MIT, can be redistributed in BSD-3 projects, but the contrary is not true. Therefore, most of our license violations presented on Table~\ref{tab:incompatible_licenses}, come from a project using MIT license that uses/includes licenses that are not compatible regarding distribution, including, for example, the BSD-3 license. This model of JavaScript software development will need support and special attention from platforms, in special, the licensing process, to avoid legal issues when composing new software using package managers.

\vspace{0.2cm}
\noindent
\textbf{License granularity: project-level vs. file-level}. Given the previous discussion on the development model introduced by JavaScript package managers, we argue that JavaScript developers are not completely aware of such a problem. More than 95\% of the respondents were somehow involved in the licensing process of a project in our survey. Yet, $\sim$63\% of them claimed that they are not aware of license incompatibility issues, and consequently, violations imposed by such behavior. This is also corroborated in our results, where $\sim$70\% of the projects use only one main license. From these, $\sim$31\% have at least two additional licenses originated from the library importing process. From our data, we can observe that the behavior established by package managers may have a direct effect on how JavaScript projects are licensed. Most of the analyzed projects have issues concerning licenses provided at the file-level, which in most cases, are incompatible. Yet, most of these issues could be solved if developers had expertise or additional support to solve license incompatibility issues. Larger projects  (e.g., \textsf{JXCore}, \textsf{Node.js}, \textsf{Kibana}) have already tackled this issue by bringing part of the file-level licenses from their libraries to the project-level, combining open source licenses and non-open source licenses, into the LICENSE file. Although there are tools to aid developers in this tasks --- a few examples are SPDX Validation Tool\footnote{\url{http://13.57.134.254/app/}}, FOSSology\footnote{\url{https://www.fossology.org/}}, findOSSLicense\footnote{\url{http://findosslicense.cs.ucy.ac.cy/}}, REUSE\footnote{\url{https://reuse.software/}} --- most of them are not applied accordingly on projects, and such tools may provide limited advice when combining open source license. This scenario is another hint that JavaScript developers should receive extra support when licensing projects that use package managers such as NPM.

\vspace{0.2cm}
\noindent
\textbf{The license of test code files}.
In addition to source code files, we also explored the license usage of test code files. This is particularly important because test code might not always be part of the software distribution. In this case, this would not create an incompatibility if the test code use a license that is incompatible with the license the project is licensed under.
We then studied 17,466 test files that have license declarations spread over the 1,552 studied projects. These files represent only 4\% of the overall analyzed files. We could only study test files with license declaration because we filtered out from our dataset files that do not have any license declaration.
We found these files by searching for ``test'' in the file name. We also made some manual investigation to make sure that the selected files are indeed test files.
From these test files, we did not find any case in which a test file was licensed under a different license than the project. The three most recurring licenses in test files are: Apache-2.0 (6,349 test files, 36\%), MIT (4,756 test files, 27\%), and BSD-3 (2,059, 12\%).
This approach has important limitations, though.
First, since we filter out files without license declaration, we may be missing many test files\footnote{\url{https://github.com/nodegit/nodegit/blob/02e617bea465ae132bffe012325547875ff73b73/test/tests/convenient_line.js}}.
Second, our approach may also miss test files that do not have ``test” in the name of the file (we noticed that some JavaScript projects only use ``test” to name their testing folders, but not their testing files\footnote{\url{https://github.com/nodegit/nodegit/tree/02e617bea465ae132bffe012325547875ff73b73/test/tests}.}
We leave an extensive investigation of licensing of test files for future work.

\vspace{0.2cm}
\noindent
\textbf{Are test files being distributed?}
Here we shed some light on whether these test files are indeed being distributed. To do so, we mined the \texttt{.gitignore} files.
Ignoring test files with the \texttt{.gitignore} file is fairly common in open source projects. Indeed, the FSFE (Free Software Foundation Europe e.V.) considers this a good practice, and explicitly list it in their FAQ\footnote{\url{https://reuse.software/faq/\#exclude-file.}}
In our dataset, we found \texttt{.gitignore} in 1,389 JavaScript Project. When analyzing these files, we noticed that 384 projects have rules that ignore test files, avoiding them to be committed into the shared repository.
Moreover, some of the relus ignore directories that keep test files, e.g., \texttt{test/*} or \texttt{tests/*}. Some projects also ignore \textsf{Mocha.JS} test files  (\texttt{mocha.js}, \texttt{/mocha-test-results.html}). \textsf{Mocha.JS} is a JavaScript testing framework, widely used in the JS community. A few projects also ignore the test outcomes (e.g. \texttt{/build/test/*}). Overall, approximately 25\% of the projects in our dataset do not have their tests committed to their repositories.

\subsection{Implications}

Our work has implications for researchers, practitioners, and instructors.

\vspace{0.2cm}
\noindent
\textbf{Researchers.} Researchers can take advantage of our work in several ways. First, we highlighted that multi-licensed projects are common in our dataset (RQ1). Researchers could take this information to investigate further the degree of multi-licensed projects in other open source communities. Second, we noticed that many projects use non-software license in the file-level (RQ2). Given this observation, researchers could build tools to assess whether a non-software license is indeed being used to license a non-software artifact. This could assure that software using proper licenses, code is not being licensed under non-software licenses.
Furthermore, there are scenarios in which license incompatibility becomes harder to identify with traditional software mining repository techniques. In this context, researchers could investigate license incompatibility issues not only if two source code files licensed under different licenses exist within a given project (RQ3) but also consider if these two files are using each other. %For instance, researchers could take advantage of type systems to make sure a given source code file is a dependency of another source code file.

\vspace{0.2cm}
\noindent
\textbf{Practitioners.} Our results could also be useful for practitioners. One of our first observations is that multi-licensed projects exist, and there are many (RQ1). Moreover, in our survey, we also observed that developers still do not fully understand the implications of using multiple licenses (RQ5). We then envision recommendation tools that could guide developers on their license decision-making. If a license inconsistency is found, such a tool could recommend alternative implementation that conforms to the type specification~\citep{Oliveira:MSR:2019}.
Moreover, we observed that multi-licensed open source projects do not consistently warn potential clients that the software is multi-licensed (RQ4).
In this scenario, we envision that social coding environments, such as GitHub or GitLab, could not solely rely on the information available on the \texttt{LICENSE} file, since we noted that some projects use one \texttt{LICENSE} file per license used. These coding environments should also inspect the entire codebase looking for additional license declarations.
We also recommend adopting practices that allow easy, automatic verification of project licenses, permissions, and copyright owners like FSFE's REUSE\footnote{\url{https://reuse.software/}} and Linux Foundation's SPDX\footnote{\url{https://spdx.dev}}.

\vspace{0.2cm}
\noindent
\textbf{Instructors.} Instructors of software engineering courses could also benefit from this work. Since we observed that 77\% of our respondents were somehow involved with licensing issues, we believe that instructors could update their teaching agenda to cover the importance of legal aspects in open source software.

\subsection{Limitations}

Like any empirical study, this one also has several limitations and threats to validity.

First, we relied on a single tool to perform our analysis: the ScanCode Toolkit. ScanCode is adopted by open source organizations like Eclipse Foundation, OpenEmbedded.org, the Free Software Foundation, RedHat, Fabric8, and many others\footnote{\url{https://directory.fsf.org/wiki/ScanCode\_Toolkit}}. To make sure that the results reported by this tool were accurate, we complemented the analysis with a manual investigation of half (\totalProjectsAnaliseManual) of the studied projects, as reported in Section~\ref{sec:validacaoManual}. We did not manually investigate all projects because it was too time-consuming (it took us one to two weeks to conduct this process). Moreover, 50\% is often far beyond sample size, with a confidence level of 95\% with a ±5\% confidence interval. When performing this analysis, we noted that about 95\% of the results reported by ScanCode were accurate, which we consider satisfactory. Finally, the results can be slightly different if a distinct tool is adopted by researchers to replicate our work.

Second, in this work, we focused only on understanding multi-license usage in popular JavaScript projects. Although we agree that this could limit the results' generalization, we believe that this population is particularly appropriated for this work. The reason is two-fold: First because the JavaScript programming language is facing an unprecedented growth in popularity, and several JavaScript libraries are also extremely popular among developers. Second, because JavaScript developers could reuse these libraries by merely including one single (often minified) source code file in their codebase. This high degree of popularity associated with an easy to reuse could bring challenges to license understanding and usage. Additionally, our results are majorly focused on the NPM package manager's projects: more than 84\% of the projects analyzed are distributed via NPM.

Third, our work relies mostly on data extracted by software repositories. We tried to mitigate this limitation by conducting a survey with developers that might be aware of license issues. To find these developers, we deployed our survey to \totalInvitedSurvey developers that had made changes to license-related files. However, we only received feedback from 6\% of them. We suspect that this low response rate is related to the roles that these developers play in the studied project. Since project maintainers might be the ones that could make decisions regarding license usage, these maintainers are also often invited to fill many academic surveys. Even though our survey was short and with several closed questions (which made it straightforward to answer), we believe that these maintainers might not be willing to answer yet another survey.

%To analyze the license information of the projects in question, we initially used a package developed in JavaScript and distributed by the NPM \footnote{NPM is the package manager for Node.js. Available in:\url{https://www.npmjs.com/}} package manager, denominated License Checker\footnote{Available in: \url{https://www.npmjs.com/package/license-checker}}, this tool analyzes the repository and returns the license(s) found in it. But the results obtained by the tool were unsatisfactory because of the 1,552 repositories analyzed by the tool, were found to license only about 300 projects (less than 20\% of the research population). This limitation exists due to license analysis tools in projects that use NPM as a dependency manager.

\section{Related Work}

%Open source software projects and their community are constantly subjects of recent research.
Many studies were conducted in the recent years regarding the use, evolution, and problems related to open source software licenses~\citep{DiPenta:ICSE:2010,Vendome:2017:ESME, 7332449,Meloca:2018:MSR,Vendome:2018:ICSE,Duan2017, Paschalides2016, 7985655}. Most of these studies are related to their use and, possible implications and violations of such use. For instance, \cite{DiPenta:ICSE:2010} proposes a method to track the evolution of software licensing and investigated its relevance on six major open source projects, presenting inconsistencies related to files without a license. License adoption and their evolution was also subject of analysis~\citep{Vendome:2017:ESME, 7332449}. Recently, \cite{Meloca:2018:MSR} presented the effects of non-approved license use on multiple package managers, including an analysis of licensing evolution in these environments.

Some authors focused on understanding license inconsistency issues. For instance, \cite{GermanMSR,Wu:2017:EMSE} studied license inconsistencies in general. Although they mention that some projects adopt a dual license model, they did not conduct a comprehensive study on the use of multiple licenses. \cite{Vendome:2018:ICSE} have recently mapped what they called ``licensing bugs'', presenting a taxonomy of licensing issues, including misinterpretation and license misuse, license violations, as well, license documentation, which analyzes how licenses are documented in a software project. The authors claim that a mismatch between the documented license and the source code licensing might stall the software distribution in certain communities, but it generally does not impact the package acceptance on package managers, as we notice when analyzing our dataset.

More related to our work, some research has been conducted regarding multi-licensing issues. Most of them are related to licensing inconsistencies and issues related when combining reusable artifacts from different sources. Several tools and methods were developed to check license inconsistencies to help developers avoid such issues~\citep{Duan2017, Paschalides2016, 7985655}.
Over the last two decades, some studies were conducted to comprehend the dual licensing model~\citep{valimaki2003dual,comino2011,koski2005oss}. These studies analyzed the dual licensing model that several large companies started to adopt from mid-2000s on, releasing their software under two licenses, in general, one proprietary or copyleft license, and the more permissive one. These works analyzed data from such companies, inspecting possible commercial sustainability scenarios, which was an uncertain at the time, legal issues when combining a copyleft OSS license, such as GPL, and copyrighted commercial software.
Other studies were directed to create models and frameworks to guide companies during this transition process to open their source code~\citep{german2009,holck2007framework}.

More recently, to understand how developers perceive the open software licensing process, including dual and multi-licensing, \cite{Almeida:2017:ICPC} surveyed 375 developers regarding the use of three popular licenses both alone and in combinations. The results show that, although understanding the use of individual licenses, the consequences and results of multiple licensing are not completely clear to developers, suggesting the use of supporting tools for the task.

To the best of our knowledge, however, there is little understanding on the use of multiple open source licenses (and the potential implications of it). The study presented in this paper is orthogonal to previous work. We aim at tackles this gap by investigating the usage of multiple open source licenses in the vivid community of JavaScript projects due to their popularity growth and the higher chance to propagate license inconsistencies~\citep{Meloca:2018:MSR}. We focused on understanding how (RQ1, RQ2, RQ4) and why (RQ5) JavaScript Projects adopt multiple open source licenses and what is its impact in terms of inconsistencies (RQ3), shedding light in this phenomena.
%Differently from most approaches, our work focus on the recent popularization of JavaScript package manager NPM. Yet, it is unclear why and how users combine packages with different licenses in this ecosystem.

\section{Concluding Remarks}

In the context of open source projects, open source software licenses drive how source code can be modified, reused, and redistributed.
Although software developers still struggle to understand license usage details, most of the open source projects available out there do have a license.
What is so far unclear is that open source projects could (and often do) employ more than one license.
Unfortunately, either social coding websites (such as GitHub) or package manager systems (such as NPM) provide little guidance on describing when a given software project uses more than one license.
This lack of awareness brings challenges to developers, since when a software project uses more than one license, clients of this project should place additional care in understanding if all the licenses used to comply with the license that the clients use in their own projects.

In this work, we shed some initial light on JavaScript projects that employ more than one license. We were particularly interested in understanding two questions: (1) how common do JavaScript projects use multiple licenses?, and (2) why does this happen?
To answer this question, we conducted a mix-method study, combining quantitative data acquired from \totalProjects software repositories and then we cross-validated with a survey with \totalSurvey developers.
Among the findings, we observed that the majority of the studied projects (62\% of them) use more than one license.
When studying the pair of licenses, that is, the group of licenses that are mostly employed together, we observed that pairs based on permissive licenses are the most common ones (such as the ``BSD-3 \& MIT'' pair, the ``Apache-2.0 \& MIT'' pair, and the and ``Apache-2.0 \& BSD-3'' pair).
Among the licenses used, the majority of them (93\%) are software licenses, while the remaining ones are non-software licenses.
From the survey and after manually analyzes 195 commits involving license-related files, we found 3 main reasons of why developers adopt multiple licenses: (i) take care of license compatibility issues, (ii) lead with legal third-party adoption, and (iii) and make existing license less restrictive.

\subsection{Future Work}

We plan to expand the scope of this work in several ways. First, we intend to cover additional programming languages as a way to compare our results.
We also intend to dig further into the eventual license violations that could emerge in these multiple licensing scenarios.
Additionally, we have plans to investigate test files included in the source code and their license usage, which might reveal other aspects of license combinations and possible, unexpected violation scenarios.
Finally, we also have plans for building tools that could help developers understand when they could safely incorporate third-party libraries in their codebase, considering their (multiple) license usage.

\setcitestyle{square}
\bibliography{mybibfile}

\end{document}